\documentclass[a4paper,11pt]{article}
\usepackage{jheppub} 
\usepackage{xcolor}
\usepackage{multirow}
\usepackage{subcaption}
\usepackage{float}
\usepackage{booktabs}

\title{Eigenvalue distribution from bootstrap estimates}

\author{Samuel Kov\'{a}\v{c}ik,}

\author{Katar\'ina Magdolenov\'a}
\affiliation{Comenius University Bratislava}

\abstract{The bootstrap method has proven useful for a wide range of matrix models. Here, we show that the computed momenta can be used to reconstruct the underlying eigenvalue probability distribution, which in turn allows us to compute the free energy of the model—a necessary quantity for identifying the thermodynamically preferred solution. We verify the method on the well-studied quartic potential model and then apply it to a recently analysed asymmetric multi-trace model. We consider an extended class of possible solutions and demonstrate that free-energy analysis reliably selects the correct one, making it an essential tool for studying models with a complex solution structure.}

\begin{document}
\maketitle

\section{Introduction}

Various physical systems can be expressed in terms of matrices of a certain type, for example, Hermitian matrices. When studying a model, for example, from a statistical or quantum mechanical point of view, integration over an ensemble is performed, which in the case of matrix models is an integration over the space of matrices. There are many approaches to performing this task; sometimes the model is approachable by (semi)analytical approaches, and sometimes one is required to perform numerical simulations, see \cite{Eynard:2015aea, Jha:2021exo} and references therein. Recently, the bootstrap method gained popularity \cite{Han:2020bkb}. While it requires some numerical steps, it seems to be able, at least in some cases, to obtain high-precision results with modest computational resources. The basic idea is that one sets up a recurrent equation for the moments of the underlying distribution of the studied field and has an requirement that the solution has to satisfy; this requirement does not have to be interesting from a physical point of view. Then, one needs to scan the space of a small number of initial momenta, use the recurrent equation to compute more of them and then check if they satisfy the set requirement or not. In a series of papers \cite{Lin:2020mme,Han:2020bkb,Khalkhali:2020jzr,Hessam:2021byc,Kazakov:2021lel,Aikawa:2021qbl,Berenstein:2021loy,Bhattacharya:2021btd,Zheng:2023bjj,Khalkhali:2023onm,Kazakov:2024ool,Lawrence:2024mnj,Huang:2025sua,Blacker_2022,Berenstein:2021dyf,Berenstein:2022ygg,Khan:2022uyz,Nakayama:2022ahr,Tchoumakov:2021mnh}, this method has been proven to work well for various (not only) matrix quantum mechanical problems. 

Matrix models often appear in the context of quantum spaces, and this is the point of view of interest of the authors. In these models -- and in many others -- one is often interested in observables that depend only on the matrix eigenvalues and not on the entire matrices. In some cases, referred to as pure potential matrix models, the entire action is in the form that depends only on the eigenvalues. This, however, does not mean that the other matrix degrees of freedom are not important, as one still needs to perform integration over all of them. The integration over all degrees of freedom can be split into the integration over the eigenvalues and over the rest of variables, and the other integration can be performed, producing an effective term in the action that can be interpreted as a logarithmic repulsion between the eigenvalues. This new term leads to interesting properties that have been studied before \cite{RevModPhys.69.731}. For the content of this paper, we need just one fact: the behaviour of pure potential matrix models is entirely captured by a single function –– the eigenvalue distribution from which everything else can be computed. For some other models, such as fuzzy space models, this function does not capture every relevant aspect, but it is still valuable for analysing the model, for example, from the perspective of its phase space structure \cite{Szabo:2001kg,Karabali:2006eg,Balachandran:2005ew,Ydri:2016dmy,Tekel:2015uza,Subjakova:2020prh}. 

The focus of this paper is on the reconstruction of the eigenvalue distribution from its moments that are accessible by the bootstrapping algorithm. For this, we use the method developed by Tekel and Cohen \cite{tekel2012constructing}. We show the usefulness of this construction by considering an asymmetric multitrace pure potential model that has been shown recently to have a coexistence of various solutions, that is, various local minima of free energy, of which only one is the thermodynamically preferred \cite{bukor2024simple}. While the bootstrap algorithm finds many (all, in some limit) of these solutions, the reconstructed eigenvalue distribution allows us to compute the free energy and find the preferred one. From this point of view, constructing the eigenvalue distribution is a crucial step in the analysis of the models with a complex structure of free energy minima.

This paper is organised as follows. First, we briefly summarise the bootstrap method and the eigenvalue density reconstruction from the moments. Then we show the method on a well-studied pure potential quartic model to show that the method recovers the known results. Finally, we apply the results to the recently studied findings of an asymmetric multitrace matrix model. 

\section{Matrix models and their moments}

We will consider a matrix ensemble \cite{Eynard:2015aea,Tekel:2015uza,Livan_2018} with the partition function of the form
\begin{equation}
\label{eq:partitionfunction}
    Z=\int e^{-N^2S(\phi)}d[\phi],
\end{equation}
where $[\phi]$ means the integration goes over all matrix elements, we will consider a set of Hermitian matrices of size $N$; other choices also appear in the literature. The degrees of freedom of the model are therefore the elements of the matrix $\phi$. The observables one is interested in are often expressed as traces of some function of $\phi$ and their mean value is taken to be
\begin{equation} \label{eq:expectedvalue}
\left\langle f \right\rangle = \frac{1}{Z}\int  \,  f(\phi)e^{-N^2S(\phi)} d[\phi].
\end{equation}
The simplest version of the action is a trace of some function of $\phi$, that is
\begin{equation}
    S(\phi) = \frac{1}{N}\sum_{n=0}^N g_n  \text{Tr}(\phi^n),
\end{equation}
however, later in the paper, we will also consider a multitrace model. In some situations, one can introduce terms with additional (usually) fixed matrices that add a physical context to the model, for example, a kinetic term of the form $\sim \text{Tr}[L_i,\phi]^2$, where $L_i$ are fixed matrices, in fuzzy space models. 

For some models, the action does depend only on the eigenvalues of $\phi$, denoted $\lambda$. However, one cannot just ignore the rest of the degrees of freedom, as integration initially went over all matrix elements, and the change of variables produces a Jacobian that can be added to the action as an additional term in the action:
\begin{equation}
    S_{J}(\lambda) = - \frac{2}{N^2} \sum \limits_{i<j}\log |\lambda_i-\lambda_j|.
\end{equation}

Togather, the total free energy is
\begin{equation} \label{eq:free_energy}
    S_{\mbox{eff}} = S(\lambda) + S_{J}(\lambda).
\end{equation}

An important set of observables is the moments of the eigenvalue probability distribution
\begin{equation} \label{eq:moments}
m'_n= \frac{1}{N} \left\langle \text{Tr}(\phi^{n}) \right\rangle=\int_{C} \lambda^n\rho(\lambda)d\lambda ,
\end{equation}
where $\lambda$ denotes the eigenvalues of $\phi$ and $\rho(\lambda)$ is its probability distribution with respect to \eqref{eq:partitionfunction} and $C$ denotes the support of the distribution.

The recurrent relation between the moments is given by the Dyson-Schwinger equation
\begin{equation} \label{eq:derivacia}
\int \, \frac{\partial}{\partial \phi_{ij}} \left( (\phi^k)_{ij} e^{-N^2S(\phi)} \right)  d[\phi] = 0,
\end{equation}
which for pure-potential single-trace actions, $ S(\phi) = \frac{1}{N}\text{Tr} \, V(\phi)$, yields
\begin{equation}\label{eq:generalloop}
N\left\langle \text{Tr}(\phi^{k} V'(\phi)) \right\rangle=\sum_{q=0}^{k-1} \left\langle \text{Tr}(\phi^q) \right\rangle \left\langle \text{Tr}(\phi^{k-1-q}) \right\rangle.
\end{equation}
Depending on the particular form of the action, some of the moments can be expected to vanish. The first one is a normalisation factor which is set, $m'_1=1$, the rest is then computed using the equation \eqref{eq:generalloop}.

\begin{figure}[H]
    \centering
    \begin{subfigure}[b]{0.45\textwidth}
        \centering
        \includegraphics[width=\textwidth]{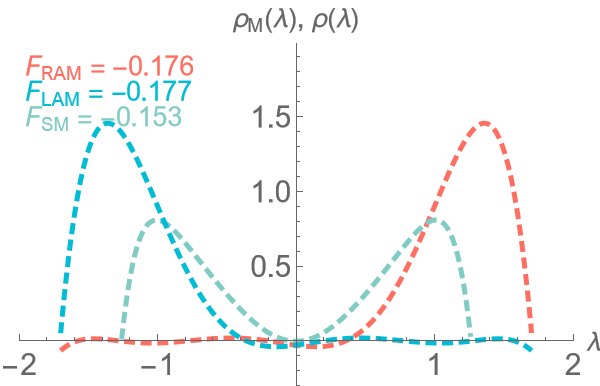}
        \caption{r=-5.265 g=1.732525}
        \label{fig:sub1}
    \end{subfigure}
    \hfill
    \begin{subfigure}[b]{0.45\textwidth}
        \centering
        \includegraphics[width=\textwidth]{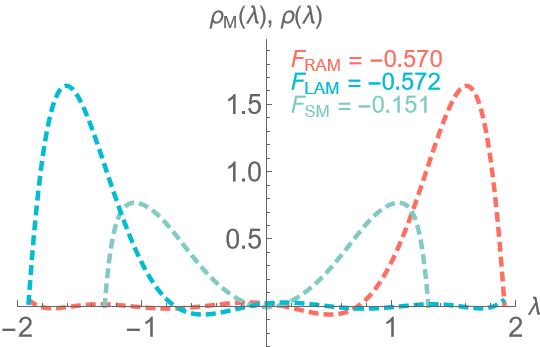}
        \caption{r=-5 g=1.5}
        \label{fig:sub2}
        
    \end{subfigure}
    \caption[Free energy for specific values r and g –– the asymmetric model]{Comparison of the free energies: \( F_{\text{RAM}} \) (right asymmetric solution), \( F_{\text{LAM}} \) (left asymmetric solution), and \( F_{\text{SM}} \) (symmetric solution) for given values of \( r \) and \( g \).}
    \label{fig:freeenergytriplepoint}
\end{figure}

The second step is to find a constraint that can be used to discard the inconsistent sets of moments $m'_i$. Usually, in the context of quantum-mechanical matrix models, one can devise a set of positive observables \cite{Han:2020bkb}. Here, we use a different method that is closer to what we seek to construct later on, that is, the underlying eigenvalue distribution. The  Hamburger problem states that there exists a sequence of moments $m'_n$ such that, for a positive Borel measure $\mu$, the following relation holds:
\begin{equation} \label{eq:momenty}
m'_n = \int\limits_{-\infty}^{\infty} x^n \, \mathrm{d}\mu.
\end{equation}
This condition is satisfied if and only if the matrix constructed from these moments $m'_n$ is positive semidefinite:
\begin{equation} \label{eq:semidefinitnost}
 M =
\begin{pmatrix}
m'_0 & m'_1 & m'_2 & \dots \\
m'_1 & m'_2 & m'_3 & \dots \\
m'_2 & m'_3 & m'_4 & \dots \\
\vdots & \vdots & \vdots & \ddots
\end{pmatrix}   
\succcurlyeq 0.
\end{equation}
This matrix is known as the \emph{Hankel matrix}, and the semidefiniteness condition can equivalently be expressed as:
\begin{equation} \label{eq:semidefinitnost_alternative}
\sum_{i,j \geq 0} m'_{i+j} c_i c_j \geq 0,
\end{equation}
for all sequences $\{c_i\}$ with only finitely many nonzero elements. By computing determinants of various submatrices of \eqref{fig:freeenergytriplepoint}, one obtains various constraints against which the obtained set of moments can be tested –– and usually discarded. 

Having both the recurrence relation \eqref{eq:generalloop} and the constraint from \eqref{eq:semidefinitnost}, we need to specify which of the moments will be considered initial and which are going to be computed; the number depends on the order of the recurrence relation. One then scans over the space of initial values of the moments and verifies which values of initial moments a set of consistent, that is, constraint-satisfying, set. 

\section{Reconstructing the probability distribution}

The Hamburger problem that we have mentioned already ponders the existence of a probability distribution given a set of moments. The goal is to have a method that takes an initial probability distribution with known moments and deforms it to a different distribution with a desired moments. Here, we follow the method to construct it and investigate some details relevant to the application of matrix models under current consideration \cite{szego75, tekel2012constructing}. To begin with, let us define a set of orthonormal functions
\begin{equation}
u_n(x) = \frac{1}{\sqrt{N_n}} \sqrt{w(x)} L_n(x),
\end{equation}
satisfying 
\begin{equation}
\int u_n(x)u_m(x)\,dx = \delta_{nm},
\end{equation}
where $L_n(x)$ are polynomials orthogonal under the weight $w(x)$
\begin{equation}
\int w(x)L_n(x)L_m(x)\,dx = N_n \delta_{nm}.
\end{equation}
Here, $N_n$ is a normalisation constant. These polynomials serve as a basis for functions on some interval
\begin{equation}
f(x) = \sum _{n=0}^{\infty} c_n u_n(x),c_n = \int f(x) u_n(x)\,dx.
\end{equation}
The moments corresponding to the weigh $w(x)$ are
\begin{equation} \label{eq:moments}
m_n = \int x^n w(x) \, dx.
\end{equation}
Now, given a set of of moments ${m_0, ..., m_n}$ we can construct a set of orthogonal polynomials 
\begin{equation} \label{eq:Kn}
K_n(x) =
\begin{vmatrix}
\renewcommand{\arraystretch}{0.8} 
\setlength{\arraycolsep}{2pt}     
\begin{array}{ccccc}
m_0 & m_1 & m_2 & \cdots & m_n \\
m_1 & m_2 & m_3 & \cdots & m_{n+1} \\
m_2 & m_3 & m_4 & \cdots & m_{n+2} \\
\vdots & \vdots & \vdots & \ddots & \vdots \\
m_{n-1} & m_n & m_{n+1} & \cdots & m_{2n-1} \\
1 & x & x^2 & \cdots & x^n
\end{array}
\end{vmatrix},
\end{equation}
that satisfy
\begin{equation}
\int w(x) K_n^*(x) K_m(x) \, dx = N_n \delta_{nm},
\end{equation}
where the normalization factor \( N_n \) is:
\begin{equation} \label{eq:normalization}
N_n =
\begin{vmatrix}
\renewcommand{\arraystretch}{0.8} 
\setlength{\arraycolsep}{0pt}     
\begin{array}{ccccc}
m_0 & m_1  & \cdots & m_{n-1} \\
m_1 & m_2  & \cdots & m_n \\
m_2 & m_3  & \cdots & m_{n+1} \\
\vdots &  \vdots & \ddots & \vdots \\
m_{n-1} & m_n  & \cdots & m_{2n-2}
\end{array}
\end{vmatrix}  
\begin{vmatrix}
\renewcommand{\arraystretch}{0.8} 
\setlength{\arraycolsep}{0pt}     
\begin{array}{ccccc}
m_0 & m_1 & \cdots & m_n \\
m_1 & m_2 & \cdots & m_{n+1} \\
m_2 & m_3 & \cdots & m_{n+2} \\
\vdots & \vdots & \ddots & \vdots \\
m_{n-1} & m_n & \cdots & m_{2n-1} \\
m_n & m_{n+1} & \cdots & m_{2n}
\end{array}
\end{vmatrix}.
\end{equation}
Note the fact that the normalisation factor $N_n$ requires first $2n$ moments. Now any probability distribution can be expanded this way
\begin{equation}
P(x) = \sqrt{w(x)} \sum_{n=0}^{\infty} c_n u_n(x),\label{eq:PD definition0}
\end{equation}
where the factor $\sqrt{w(x)}$ makes it so that the coefficients $c_n$ depend only on the moments and one can express, that is ${c_n =\frac{1}{\sqrt{N_n}}\langle L_n(x) \rangle}$, where $\langle \cdot \rangle$ is obtained with respect to $P(x)$. Note that since $L_n$ are polynomials, $\langle L_n(x) \rangle$ can be easily expressed in terms of moments of $L_n(x)$. As a result, we have 
\begin{equation} \label{eq:distribucia}
P(x) = w(x) \sum_{n=0}^{\infty} \frac{1}{N_n} \langle L_n(x) \rangle L_n(x).
\end{equation}
We can now consider ${w(x)=P(x)}$. Also, we can take another distribution $P'(x)$ with moments $m'_n$ under this weight and expand it in the same way
\begin{equation}
P'(x) = P(x) \sum_{n=0}^{\infty} \frac{1}{N_n} \langle K_n(x) \rangle' K_n(x),
\label{eq:aproximacia}
\end{equation}
where 
\begin{equation}
\langle K_n(x) \rangle' =
\begin{vmatrix}
\renewcommand{\arraystretch}{0.8} 
\setlength{\arraycolsep}{2pt}     
\begin{array}{ccccc}
m_0 & m_1 & m_2 & \cdots & m_n \\
m_1 & m_2 & m_3 & \cdots & m_{n+1} \\
m_2 & m_3 & m_4 & \cdots & m_{n+2} \\
\vdots & \vdots & \vdots & \ddots & \vdots \\
m_{n-1} & m_n & m_{n+1} & \cdots & m_{2n-1} \\
\int P'(x)dx& \int x P'(x)dx & \int x^2 P'(x)dx & \cdots & \int x^n P'(x)dx
\end{array}
\end{vmatrix}.
\end{equation}
Combining all of this, we obtain an expression for the distribution \( P'(x) \) 
\begin{equation} \label{eq:distribution}
P'(x)= P(x) \sum_{n=0}^{\infty} \frac{1}{N_n} 
\begin{vmatrix}
\renewcommand{\arraystretch}{0.8} 
\setlength{\arraycolsep}{2pt}     
\begin{array}{ccccc}
1 & m_1 & m_2 & \cdots & m_n \\
m_1 & m_2 & m_3 & \cdots & m_{n+1} \\
m_2 & m_3 & m_4 & \cdots & m_{n+2} \\
\vdots & \vdots & \vdots & \ddots & \vdots \\
m_{n-1} & m_n & m_{n+1} & \cdots & m_{2n-1} \\
1 & m'_1 & m'_2 & \cdots & m'_n 
\end{array}
\end{vmatrix} 
\begin{vmatrix}
\renewcommand{\arraystretch}{0.8} 
\setlength{\arraycolsep}{2pt}     
\begin{array}{ccccc}
m_0 & m_1 & m_2 & \cdots & m_n \\
m_1 & m_2 & m_3 & \cdots & m_{n+1} \\
m_2 & m_3 & m_4 & \cdots & m_{n+2} \\
\vdots & \vdots & \vdots & \ddots & \vdots \\
m_{n-1} & m_n & m_{n+1} & \cdots & m_{2n-1} \\
1 & x & x^2 & \cdots & x^n
\end{array}
\end{vmatrix}.
\end{equation}
While looking cumbersome, this allows us to construct the probability distribution $P'(x)$ with given moments $m'_n$ starting from an initial distribution $P(x)$ with computable moments $m_n$. While given a sufficient number of moments, any initial distribution will converge to the one with desired moments; there are choices that make this procedure more efficient –– we will discuss those shortly.

\section{Quartic model}
Here we consider perhaps the simplest nontrivial matrix model, that is, the one with a quartic potential. We begin with reconstructing the work of \cite{Lin:2020mme} and extended by considering asymmetric distributions. The action is
\begin{equation} \label{quartic_potential}
    S = \frac{1}{N}  \text{Tr}\left( \frac{1}{2}r \,(\phi^2) + g \, (\phi^4) \right).
\end{equation}
The equation that allows us to connect different moments, \eqref{eq:derivacia}, takes the form
\begin{equation} \label{eq:rekurentquartic}
\sum_{q=0}^{s-1} \left\langle \text{Tr}(\phi^q) \right\rangle \left\langle \text{Tr}(\phi^{s-1-q}) \right\rangle - Nr \left\langle \text{Tr}(\phi^{s+1}) \right\rangle - 4gN \left\langle \text{Tr}(\phi^{s+3}) \right\rangle = 0,
\end{equation}
which yields the following equation for the moments
\begin{equation}
\small
m'_{s+3} = \frac{\sum\limits_{q=0}^{s-1} m'_q m'_{s-1-q} - r m'_{s+1}}{4g}.
\end{equation}
We keep the prime symbol here to denote the known moments of an unknown distribution. The model has a known solution, and the eigenvalue distribution is either one-cut or two-cut, in both cases symmetric. The single-cut is 
\begin{equation} \label{eq:onecut}
\rho_1(\lambda) := \frac{1}{2\pi} \left( r + 2gk + 4g\lambda^2 \right) \sqrt{k - \lambda^2},
\end{equation}
for the interval \( \lambda^2 < k \), where \( k = \frac{1}{6g} \left( \sqrt{r^2 + 48g} - r \right) \) and the two-cut solution is
\begin{equation} \label{eq:twocut}
\rho_2(\lambda) := \frac{2g|\lambda|}{\pi} \sqrt{a^2 - (b - \lambda^2)^2},
\end{equation}
for the interval \( (-\sqrt{b+a}, -\sqrt{b-a}) \cup (\sqrt{b-a}, \sqrt{b+a}) \), where \( a = \sqrt{\frac{1}{g}} \) and \( b = -\frac{r}{4g} \). The model switches between those at the critical coupling $r=-4 \sqrt{g}$. We will now reproduce this result using the bootstrap method. First, from \eqref{eq:rekurentquartic} we obtain for this case the matrix \eqref{eq:semidefinitnost} is the following
\begin{equation}
M =
\begin{pmatrix}
1 & 0 & m'_2 & \dots \\
0 & m'_2 & 0 & \dots \\
m'_2 & 0 & \frac{1 - r m'_2}{4g} & \dots \\
\vdots & \vdots & \vdots & \ddots
\end{pmatrix}   
\geq 0.
\end{equation}
Note that we put $m'_1=0$ for obvious reasons, and then the entire matrix depends only on $m_2'$. The task is then to test different plausible values of $m_2'$ and see for which this matrix remains semidefinite. 

\begin{figure}[H]
    \centering
    \begin{minipage}{0.85\textwidth}
        \centering
        \includegraphics[width=\textwidth]{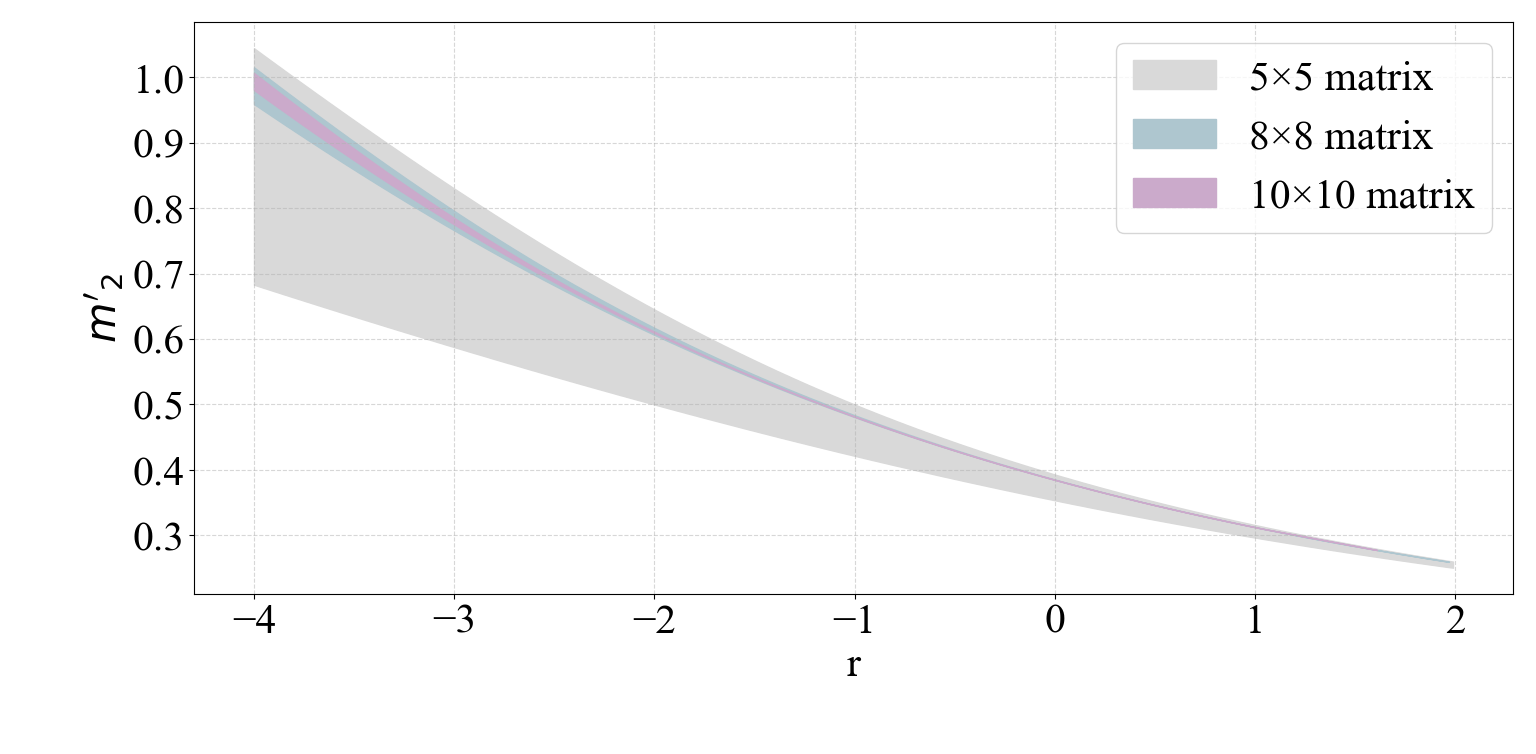}
    \end{minipage}%

    \caption [Dependence of the second moment on the constant \( r \) –– the quartic model]
    {Dependence of the second moment on the constant \( r \). To determine specific values of \( m'_2 \), it is necessary to consider larger matrices. }
    \label{fig:BMm2r}
\end{figure}

Fig.~\ref{fig:BMm2r} further suggests that the choice of \( r \) significantly influences the width of these intervals, and this dependency must be considered: for smaller values of \( r \), larger matrices were required, whereas for larger \( r \), even \(5\times5\) matrices were sufficient. The Table \ref{tab:results} shows the comparison of the bootstrap estimates and exact values of the eigenvalue distribution for different values of parameters.
\begin{table}[H]
    \centering
    \scriptsize 
    \renewcommand{\arraystretch}{1.2} 
    \setlength{\tabcolsep}{3.5pt} 
    \resizebox{0.55\textwidth}{!}{ 
    \begin{tabular}{|c|c|c|c|}
        \hline
        \multicolumn{4}{|c|}{\textbf{Exact / BM, }g=1} \\
        \hline
        $m'_n$ & \textbf{r = 1} & \textbf{r = -4} & \textbf{r = -7} \\
        \hline
        $m'_0$  & 1.0000 / 1.0000  & 1.0000 / 1.0000  & 1.0000 / 1.0000  \\
        \hline
        $m'_2$  & 0.3125 / 0.3125  & 1.0000 / 0.9998  & 1.7500 / 1.7499  \\
        \hline
        $m'_4$  & 0.1719 / 0.1719  & 1.2500 / 1.2498  & 3.3125 / 3.3123  \\
        \hline
        $m'_6$  & 0.1133 / 0.1133  & 1.7500 / 1.7497  & 6.6719 / 6.6715  \\
        \hline
        $m'_8$  & 0.0820 / 0.0820  & 2.6250 / 2.6245  & 14.0977 / 14.0969  \\
       \hline 
        $m'_{10}$ & 0.0630 / 0.0630  & 4.1250 / 4.1241  & 30.9050 / 30.9034  \\
        \hline
        $m'_{12}$ & 0.0504 / 0.0504  & 6.7031 / 6.7016  & 69.7141 / 69.7095  \\
        \hline
        $m'_{14}$ & 0.0415 / 0.0415  & 11.1719 / 11.1690  & 160.838 / 160.8260  \\
        \hline
    \end{tabular}
    } 
    \caption{Comparison of exact calculations and the bootstrap method for different values of \( r \) and $g=1$.}
    \label{tab:results}
\end{table}

\begin{figure}[H]
    \centering
    \begin{minipage}{0.3\textwidth}
        \centering
        \includegraphics[width=\textwidth]{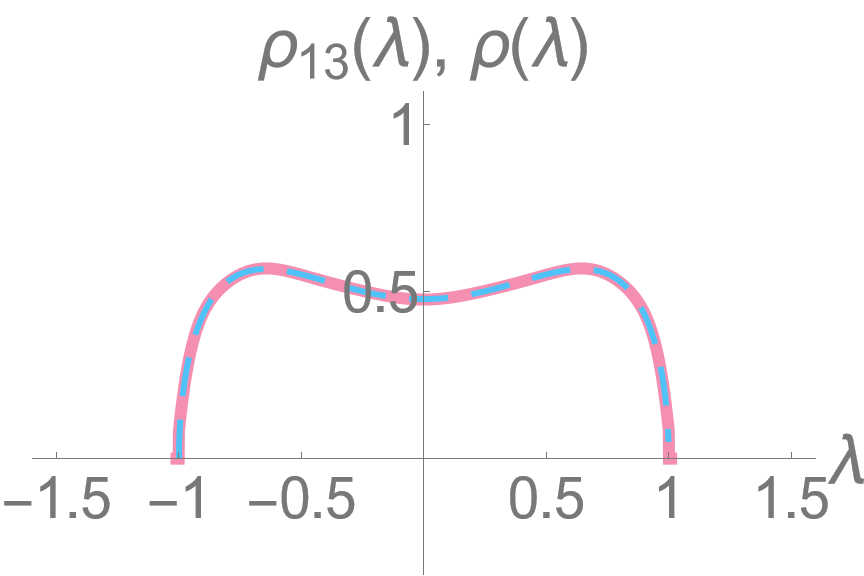}
    \end{minipage}%
    \hspace{0.2cm}
    \begin{minipage}{0.3\textwidth}
        \centering
        \includegraphics[width=\textwidth]{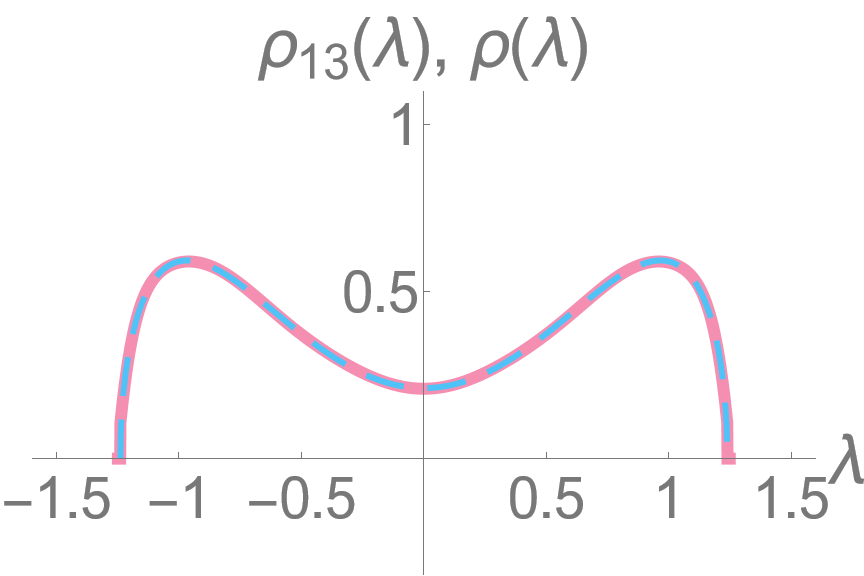}
    \end{minipage}%
    \hspace{0.2cm}
    \begin{minipage}{0.3\textwidth}
        \centering
        \includegraphics[width=\textwidth]{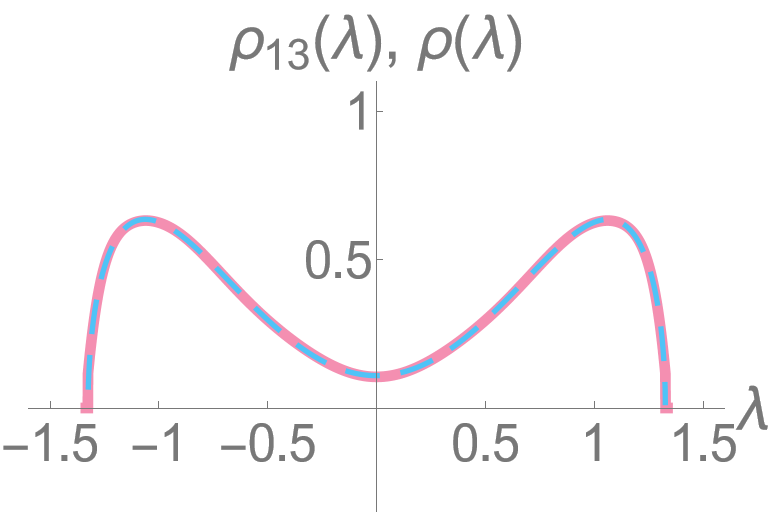}
    \end{minipage}

    \vspace{0.2cm}

    \begin{minipage}{0.3\textwidth}
        \centering
        \includegraphics[width=\textwidth]{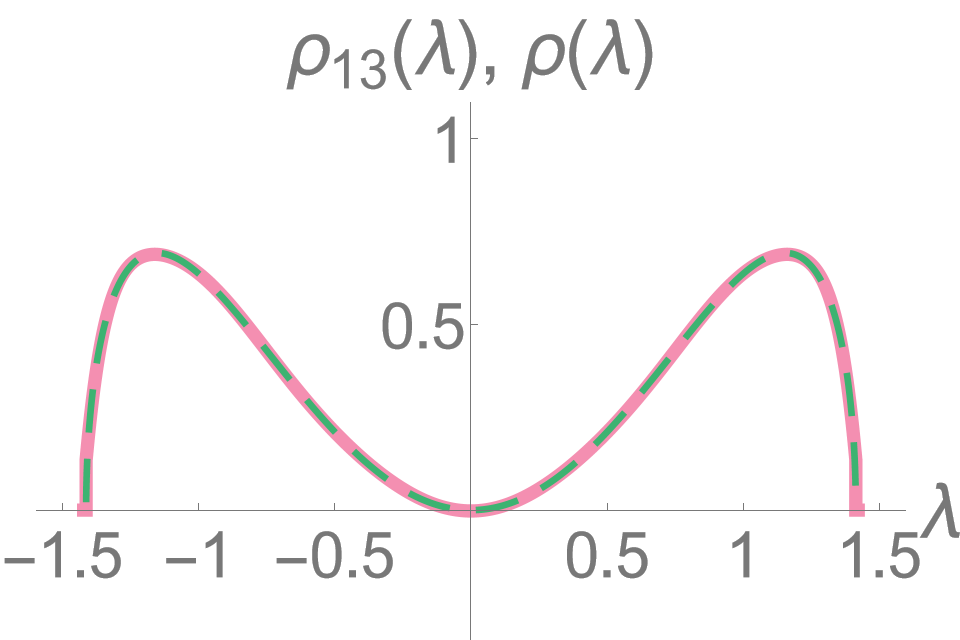}
    \end{minipage}%
    \hspace{0.2cm}
    \begin{minipage}{0.3\textwidth}
        \centering
        \includegraphics[width=\textwidth]{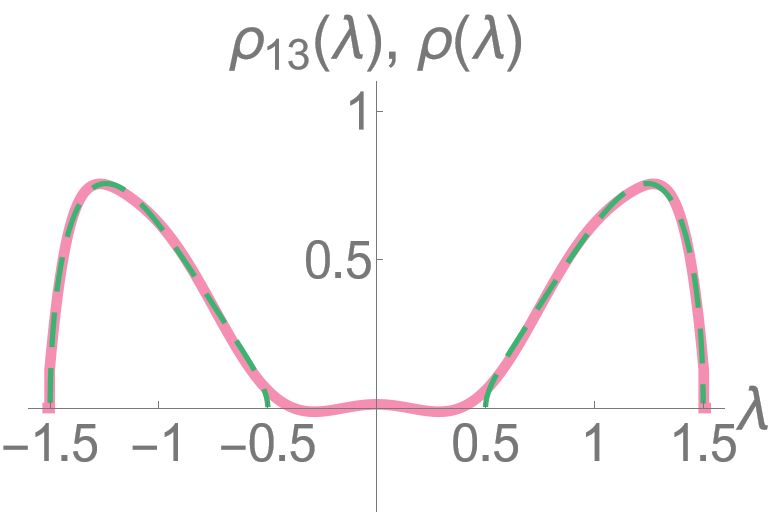}
    \end{minipage}%
    \hspace{0.2cm}
    \begin{minipage}{0.3\textwidth}
        \centering
        \includegraphics[width=\textwidth]{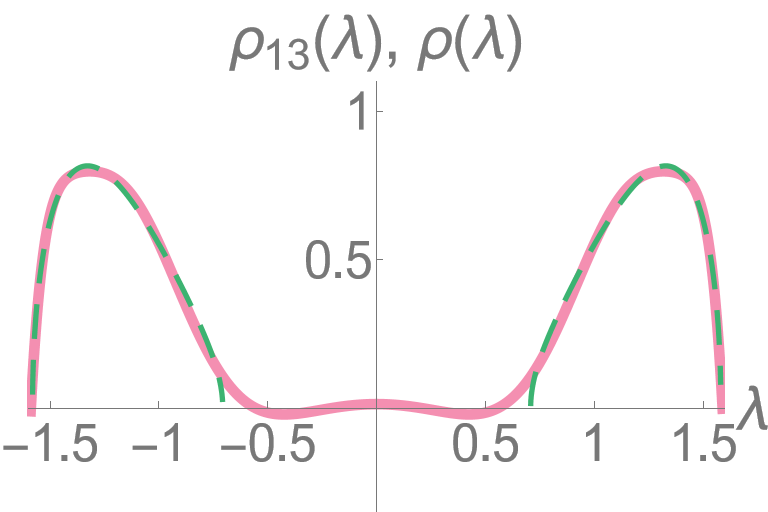}
    \end{minipage}

    \caption[Bootstrap and approximation method on the model with the quartic potential]{
        Bootstrap-generated probability densities for \( g = 1 \). The approximate distribution \( \rho_M(\lambda) \) (red solid line) is shown for \( M = 10,13 \).  
        The one-cut solution is plotted as a blue dashed line, and the two-cut solution as a green dashed line. 
        From upper left to bottom right, the plots correspond to \( r = \{1, -2, -3, -4, -5, -6\} \).
    }
    \label{fig:quarticBM}
\end{figure}

Once a suitable set of moments is obtained, we proceed with the method described in the previous section to construct the eigenvalue distribution; the examples where it is one-cut and two-cut are shown in Fig. \ref{fig:quarticBM}. Construction of the eigenvalue distribution is the novel result of this paper and as will be later showed, it is necessary for choosing the thermodynamically preferred solution.

\begin{figure}[H]
    \centering
    \begin{minipage}{0.8\textwidth}
        \centering
        \includegraphics[width=\textwidth]{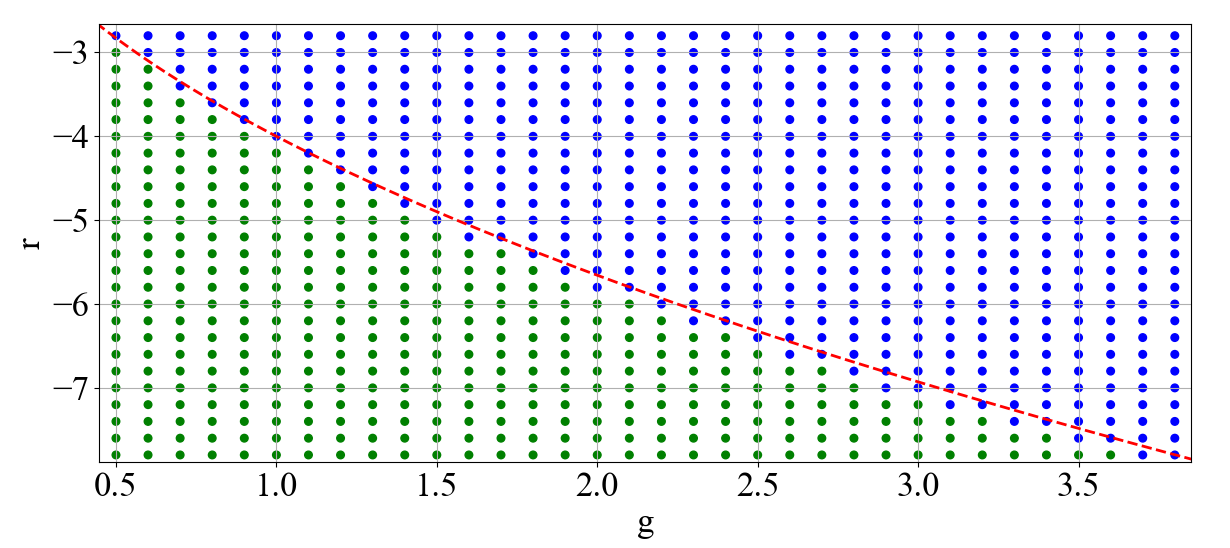}
    \end{minipage}%

    \caption[Phase diagram for symmetric solutions –– the model with the quartic potential]{ Phase diagram illustrating the transition between the symmetric one-cut (blue dots) and two-cut (green dots) solutions. The red curve \( r = -4\sqrt{g} \) marks the boundary separating the two phases.}
    \label{fig:phasediagram}
\end{figure}

Importantly, the pace of convergence depends on the particular form of the initial distribution and its support. We have tested various options, specifically: constant, Gaussian, and parabolic, and found that constant works the best for its purpose. This choice will have to be revisited for the case of asymmetric distributions. By increasing the number of moments used, for each value of the parameters $r, g$, the interval of allowed $m'_2$ quickly shrinks, and the solution quickly converges. By repeating the same process for various values of the model parameters, we have identified either a one-cut or a two-cut solution (details are commented on in the Appendix \ref{sec:ImportanceoftheInitialDistribution} and \ref{sec:NumericalDetails}) and produced the phase diagram, which is shown in Fig. \ref{fig:phasediagram}. 

\begin{figure}[H]
    \begin{minipage}{\textwidth}
        \centering
        \includegraphics[width=1\textwidth]{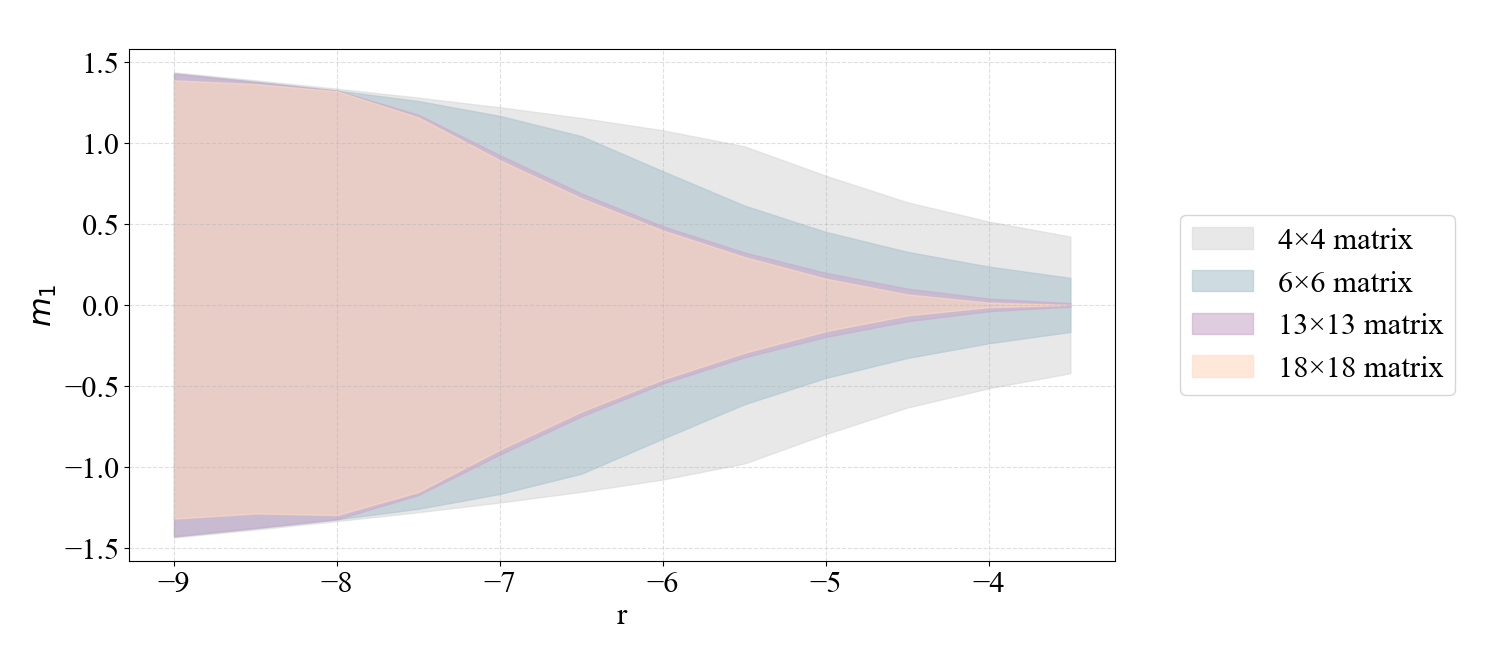}
    \end{minipage}
    \caption[Dependence of the first moment $m'_1$ on the constant $r$ –– the quartic model]{Dependence of the first moment \( m'_1 \) on the constant \( r \) for \( g = 1 \). 
    For \( r > -4 \) the values converge only to the symmetric solution (\( m'_1=0 \)). 
    For smaller \( r \), however, progressively wider symmetric intervals \((-m'_1,m'_1)\) appear, indicating a richer structure of possible first moments than in the symmetric case.}
    \label{fig:m1QP}
\end{figure}

This has all been done assuming the resulting solution is symmetric, as has already been done in the aforementioned reference \cite{Lin:2020mme}. Omitted there was the option of asymmetric distributions. We consider it by not fixing $m_1'=0$ and making it the parameter from which we start the recurrent relation, and whose values we scan over to find possible solutions. Contrary to the case of a symmetric solution, for some choices of parameters $r,g$, the set of allowed values of $m_1'$ does not converge to a single value; instead, we obtain an interval, see Fig. \ref{fig:m1QP}.
For $r<-2\sqrt 15g$, taking the endpoints of the interval, we obtain a single-cut asymmetric solution matching the analytic solution
\begin{equation} \label{eq:one-cut asymetric}
\begin{split}
\rho_3(\lambda)&=\frac{1}{2\pi}\left( 4D^2g + 4Dg\lambda + 2\delta g + r + 4g\lambda^2  \right)\sqrt{(D+\sqrt{\delta}-\lambda)(\lambda-D+\sqrt{\delta})} \\
&=\frac{1}{2\pi}\left( 4D^2g + 4Dg\lambda + 2\delta g + r + 4g\lambda^2  \right)\sqrt{\delta-(\lambda-D)^2},
\end{split}
\end{equation}
for the inteval $(D-\sqrt{\delta}, D+\sqrt{\delta})$, where 
\begin{equation}
    \delta=\frac{-r-\sqrt{-60g+r^2}}{15g} ,
    D= \pm \sqrt{\frac{-3r+2\sqrt{-60g+r^2}}{20g}}.
\end{equation}

\begin{figure}[H]
    \centering
    \begin{minipage}{0.3\textwidth}
        \centering
        \includegraphics[width=\textwidth]{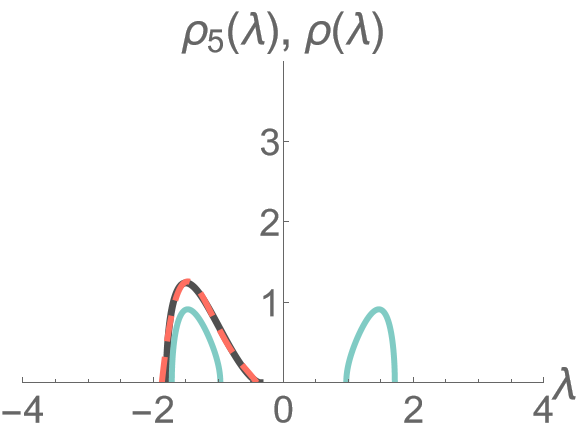}
    \end{minipage}%
    \hspace{0.2cm}
    \begin{minipage}{0.3\textwidth}
        \centering
        \includegraphics[width=\textwidth]{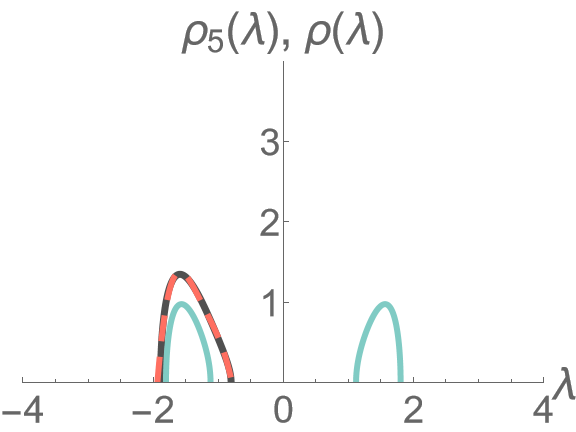}
    \end{minipage}
    \hspace{0.2cm}
    \begin{minipage}{0.3\textwidth}
        \centering
        \includegraphics[width=\textwidth]{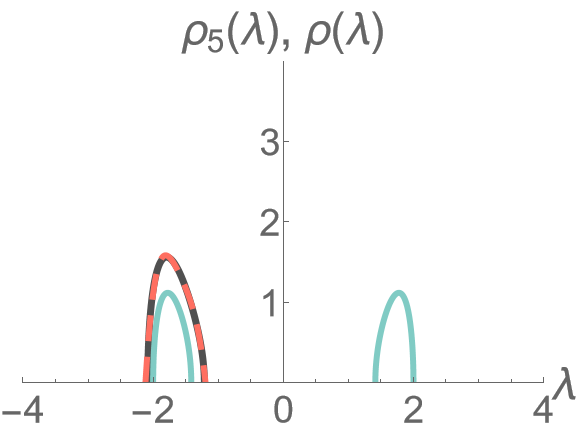}
    \end{minipage}
    \\
    \begin{minipage}{0.3\textwidth}
        \centering
        \includegraphics[width=\textwidth]{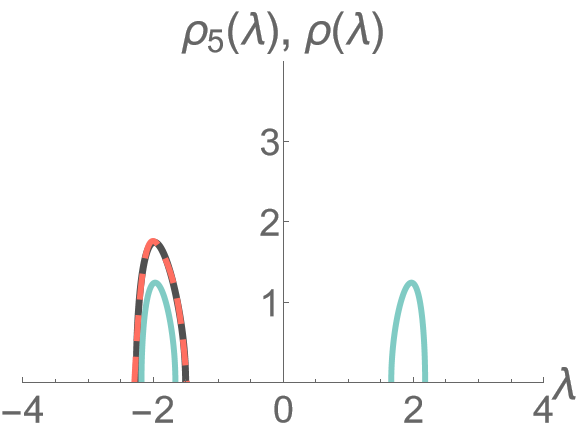}
    \end{minipage}
    \hspace{0.2cm}
    \begin{minipage}{0.3\textwidth}
        \centering
        \includegraphics[width=\textwidth]{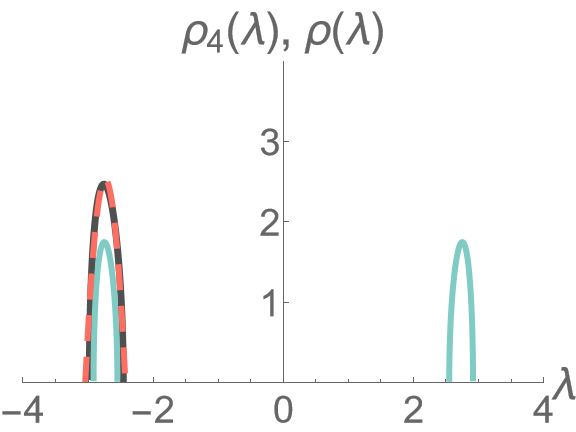}
    \end{minipage}%
    \hspace{0.2cm}
    \begin{minipage}{0.3\textwidth}
        \centering
        \includegraphics[width=\textwidth]{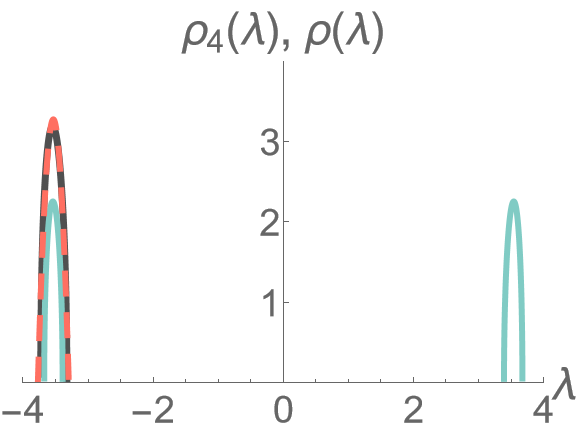}
    \end{minipage}
     \caption[Bootstrap and approximation method –– the quartic potential –– asymmetric one-cut solution]{Plot of the asymmetric one-cut solution obtained using the approximation method and moments generated via the bootstrap method (red dashed line), compared with the exact solution (black line). The green line represents the symmetric two-cut solution. The plots are shown for \( g = 1 \) and, from left to right, for \( r =\{ -2\sqrt{15}, -9, -12, -15, -30 ,-50\}\).}
     \label{fig:ASQP}
\end{figure}

For values between those, when reconstructing the probability distribution for values in between, we obtain an asymmetric two-cut solution. To choose the correct solution among the symmetric and asymmetric solutions, we compute the free energy \eqref{eq:free_energy}. The results are shown in the Table \ref{tab:free_energy}. 
\begin{table}[H]
    \centering
    \small 
    \renewcommand{\arraystretch}{1.15} 
    \setlength{\tabcolsep}{5pt} 
    \resizebox{0.6\textwidth}{!}{
    \begin{tabular}{|c|c|c|c|}
        \hline
        \multicolumn{4}{|c|}{Free energy values for $g=1$} \\
        \hline
        $r$    & Symmetric two-cut & Asymmetric two-cut & Asymmetric one-cut \\
        \hline
        $-8$   & $-3.233$ & $-1.932$ & $-1.895$   \\
        \hline
        $-9$   & $-4.353$   & $-3.065$ & $-2.885$    \\
        \hline
        $-12$  & $-8.218$ & $-7.907$  & $-6.657$ \\
        \hline
        $-15$  & $-13.361$  & $-12.277$ & $-11.598$  \\
        \hline
        $-20$  & $-24.289$   & $-24.019$  & $-22.382$  \\
        \hline
    \end{tabular}}
    \caption{Free energy for selected values of $r$ at $g=1$ for symmetric two-cut, asymmetric two-cut, and asymmetric one-cut solutions. For asymmetric two-cut solutions, multiple types of solutions exist; here we show a few randomly selected examples.}
    \label{tab:free_energy}
\end{table}

From this, we can conclude that the correct preferred solution with the minimal free energy is that of the symmetric solutions, namely the one-cut and two-cut solutions, as shown in the phase diagram in Fig.\ref{fig:phasediagram}.

To summarise the bootstrap study of the quartic potential model, the method is capable of reproducing the known results, and this effort is greatly enhanced by some choices of the initial probability distribution, and computation of free energy is needed to specify the correct solution. Let us now move to a far less-known model with an additional asymmetric potential. 

\section{Asymmetric multitrace model}

\begin{figure}[H]
    \centering
    \begin{minipage}{\textwidth}
        \centering
        \includegraphics[width=1.1\textwidth]{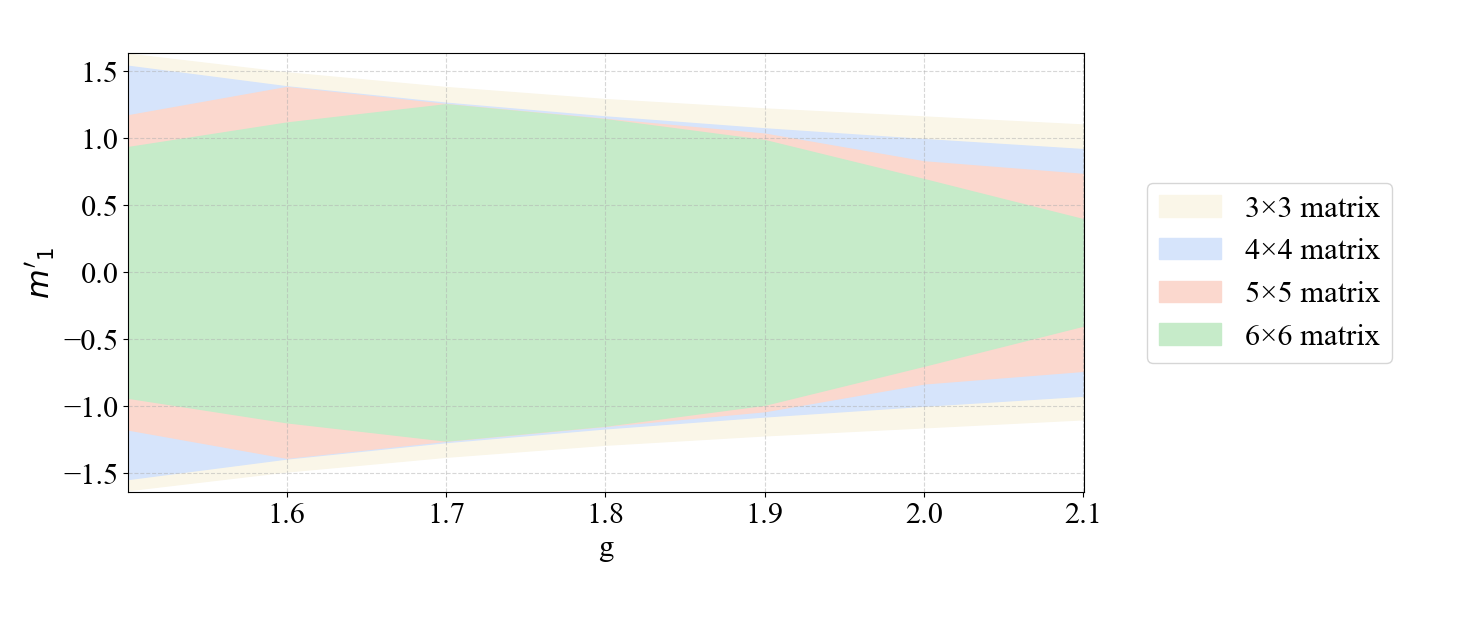}
    \end{minipage}%
    \caption[Dependence of the first moment $m'_1$ on the constant $g$ –– the asymmetric model]
    {Interval of suitable values for the first moment at \( r = -5.265 \) for various \( g \) and bootstrap matrix sizes \( M \). 
    As in the quartic case, a boundary emerges as the matrix size increases, toward which \( m'_1 \) converges. 
    The iteration step used was 0.0005; for \( g < 1.7 \) or \( g > 1.9 \), smaller steps would be needed for larger matrices to approach the limiting line.}
    \label{fig:CAMMMm1forR}
\end{figure}

Let us now consider the following model investigated in \cite{bukor2024simple}:
\begin{equation} \label{eq:asymAction}
S = \frac{1}{N} \text{Tr} \left( \frac{1}{2}r \, \phi^2 + g \, \phi^4 \right) + \frac{1}{N^2} t\, \text{Tr}(\phi)\text{Tr}(\phi^3).
\end{equation}
The Dyson-Schwinger equations~\eqref{eq:derivacia} gives the following relation between the moments
\begin{equation} \label{eq:rekurentasymmetric}
\begin{split}
& \sum_{q=0}^{k-1} \left\langle \text{Tr}(\phi^q) \right\rangle \left\langle \text{Tr}(\phi^{k-1-q}) \right\rangle - N r \left\langle \text{Tr}(\phi^{k+1}) \right\rangle  - N4g \left\langle \text{Tr}(\phi^{k+3}) \right\rangle \\
& -t \left( \left\langle \text{Tr}(\phi^{3}) \right\rangle \left\langle \text{Tr}(\phi^{k}) \right\rangle -3 \left\langle \text{Tr}(\phi) \right\rangle \left\langle \text{Tr}(\phi^{k+2}) \right\rangle \right) = 0.
\end{split}
\end{equation}
As we are dealing with asymmetric model, we cannot set $m'_1=0$ and we are left with two unspecified moments, $m'_1, m'_2$
\begin{equation} \label{eq:semidefASYM}
M =
\begin{pmatrix}
1 & m'_1  & m'_2 & \dots \\
m'_1 & m'_2 & -\dfrac{3tm'_1m'_2 + rm'_1 }{4g + t} & \dots \\
m'_2 & -\dfrac{3tm'_1m'_2 + rm'_1 }{4g + t} &  \dfrac{1-rm'_2 -4tm'_1m'_3}{4g } & \dots \\
\vdots & \vdots & \vdots & \ddots
\end{pmatrix}   
\succcurlyeq 0.
\end{equation}
Now, we can proceed as before -- scan over a two-dimensional lattice of parameters $r,g$ and find which of the values are permitted by \eqref{eq:semidefASYM}. As in the case of asymetric solutions of the quartic potential model, we find, for different values of $g$ an interval of allowed solutions that does not shrink with additional moments included, see Fig. \ref{fig:CAMMMm1forR}.

\begin{figure}[H]
 \centering
    \centering
    \begin{minipage}{0.6\textwidth}
        \centering
        \includegraphics[width=\textwidth]{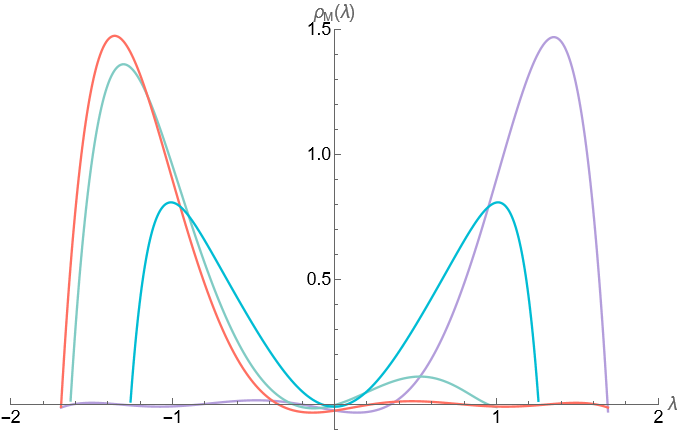}
    \end{minipage}%
 \caption[Approximate distributions \(\rho_M(\lambda)\) for \( g =  1.73252 \), \( r = -5.26502 \) --- the quartic model]{Plotted approximate distributions \( \rho_M(\lambda)\) for \( g =  1.733 \), \( r = -5.265 \). The purple and red lines represent the extreme values of the first moment, corresponding to one-cut asymmetric solutions on the right and left sides, respectively. We also observe the persistent presence of a symmetric two-cut solution (blue line), as well as several asymmetric two-cut solutions, one of which is shown as the green line.}
\label{fig:APg1.7r-5.26}
\end{figure}

Again, this is caused by various allowed asymmetric and symmetric solutions, see Figure \ref{fig:APg1.7r-5.26}. 

Therefore, to find the correct solution, we again compute the free energy and find which solution minimizes it.


Doing this for every chosen value of $r,g$, we construct the phase diagram of the model \eqref{eq:asymAction}, and we find it to be in good agreement with the one obtained in the previous study, \cite{bukor2024simple}. Note how easily the used method can be modified for a rather different form of matrix action.

\begin{figure}[H]
    \centering
    \begin{minipage}{\textwidth}
        \centering
        \includegraphics[width=1\textwidth]{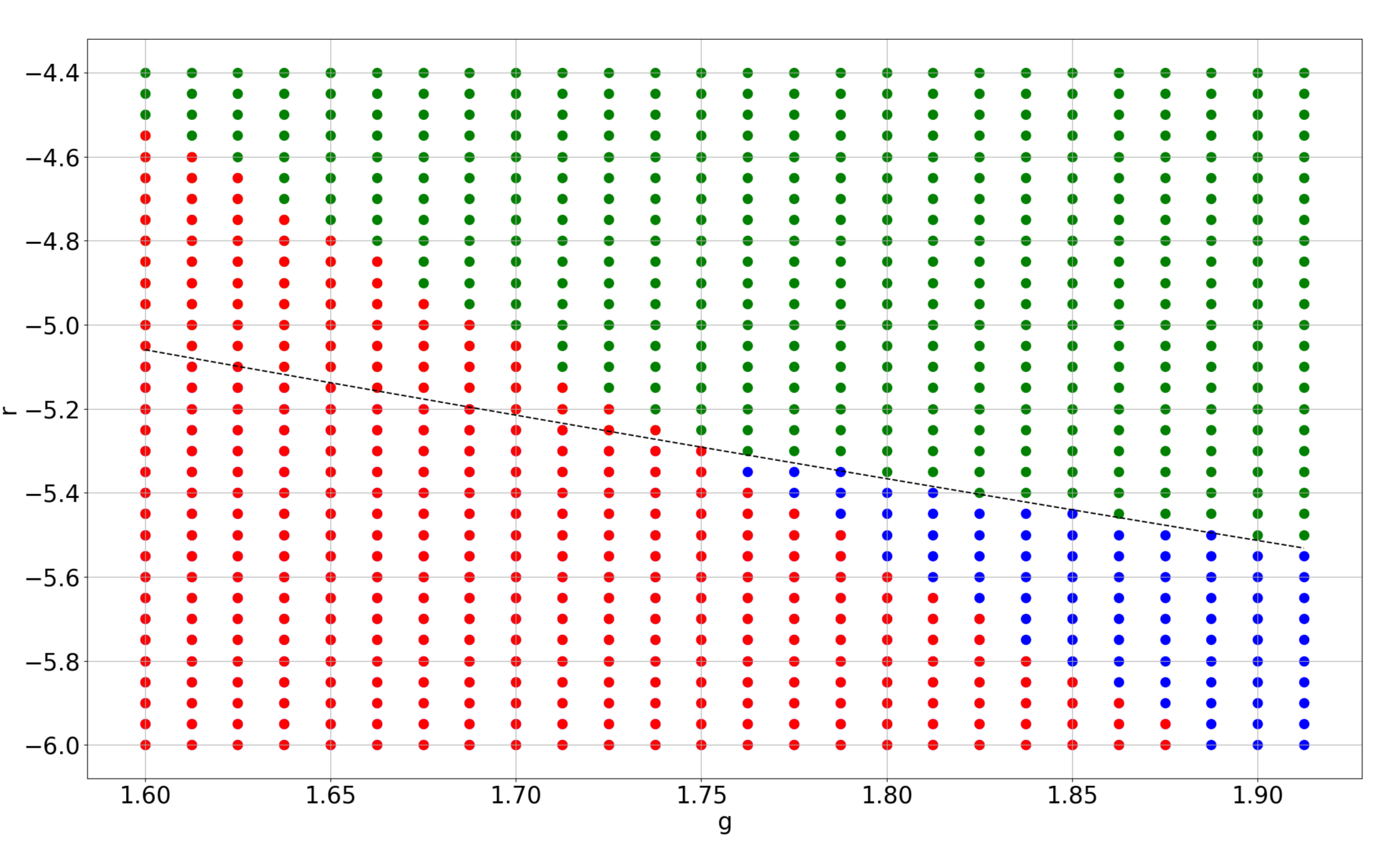}
    \end{minipage}%
\caption[Phase diagram for the cubic asymmetric multitrace matrix model]{Phase diagram for the cubic asymmetric multitrace matrix model.
Symmetric solutions are represented by blue (two-cut) and green (one-cut) dots, while the red ones correspond to asymmetric one-cut solutions. The black line indicates the transition between symmetric solutions, given by $r=-4\sqrt{g}$.
}
\label{fig:AMFDV}
\end{figure}

The following graph focuses on the region where, according to the referenced article, a triple point, where all three phases coexist (symmetric one-cut, two-cut, and asymmetric one-cut), is expected to exist.
\begin{figure}[H]
    \centering
    \begin{minipage}{\textwidth}
        \centering
        \includegraphics[width=1\textwidth]{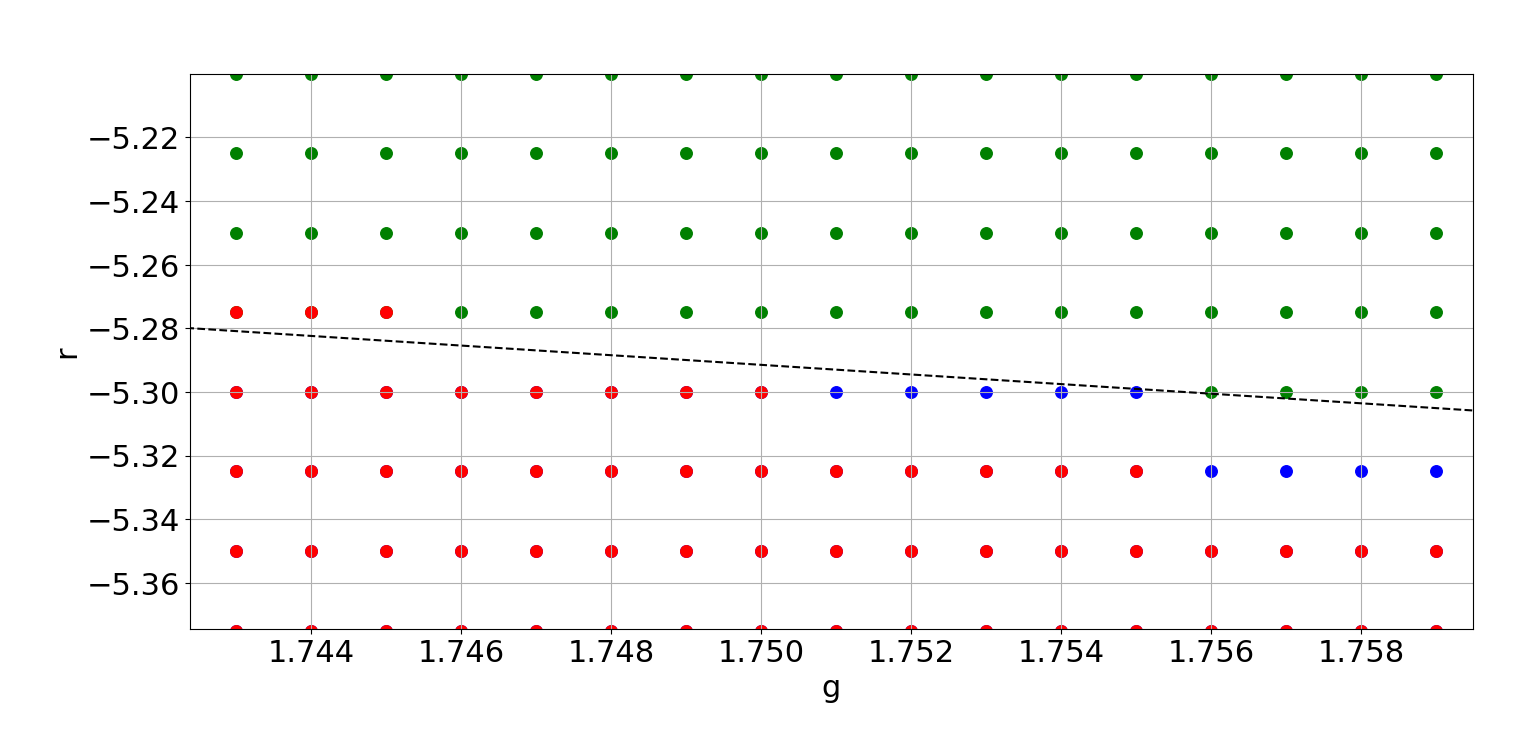}
    \end{minipage}%
\caption[Triple point for the cubic asymmetric multitrace matrix model]{ Triple point for cubic asymmetric multitrace matrix model
}
\label{fig:AMFDtriple point}
\end{figure}
Looking at the triple point in Fig.~\ref{fig:AMFDtriple point}, we can see that in our approximate approach, it appears around the point \( g = 1.75 \), \( r = -5.29 \). In the article \cite{bukor2024simple}, the triple point was found at \( g = 1.7325 \), \( r = -5.265 \), and through Monte Carlo simulations at \( g = 1.7706 \), \( r = -5.5432 \). The value obtained here is theoretically closer because the method naturally works in the large-\( N \) limit.

\section{Conclusion}

We have extended the bootstrap method for matrix models by computing the eigenvalue distribution from the moments already being used in the procedure. This allows us not only to track phase transitions but also to compute the free energy, which is necessary for selecting the correct, that is, the thermodynamically preferred, solution among the plethora found by the bootstrap method.

In more detail, we have focused on the quartic single-trace potential with a single (scalar) matrix model. This is one of the simplest and most studied examples, and we verified the method on it. Usually, the asymmetric solutions are discarded by hand by considering odd moments to be vanishing. A similar insight about the structure of solutions might be less accessible in complex models, and a method not relying on it is warranted. We have shown that without these assumptions, one also recovers asymmetric solutions which are known to exist but also to have larger free energy than the symmetric ones. 

This simple model allowed us to test the sensitivity of the result on the choice of initial probability distribution, which is being updated and transformed into the correct distribution by the method of \cite{tekel2012constructing}. We found Gaussian and constant distributions to work reasonably well. However, considering asymmetric distributions required us to choose the interval of initial distribution more delicately, which made finding the distribution more difficult. In practice, a good approximation was obtained using 10-15 moments. Note that computing the moments is simple, checking the positivity is less so. However, the total computation time relies heavily on the size of the scanned parameter space, which can be reduced by various assumptions, such as the discussed parity of the solution. Contrary to this, the usual numerical approach of Hamiltonian Monte Carlo allows for fewer such simplifications \cite{Jha:2021exo}.

We have utilised a particular form of the recurrent equation, the constraint and the set of orthogonal polynomials –– each of those could have been chosen differently and should be, if a it is proffered in current study. We were searching for a distribution on a real line, for a bounded interval one has the Hausdorff moment problem, for semibounded interval the Stieltjes moment problem or trigonometric moment problem for a unit circle \cite{simon1999classicalmomentproblemselfadjoint}.  

We have then used our findings to study an asymmetric multi-trace potential model recently analysed in \cite{bukor2024simple}. We were able to produce the phase diagram of the model by computing the free energy using the reconstructed eigenvalue distribution. 

To summarise, obtaining the eigenvalue distribution is a vital part of the bootstrap procedure, and we have shown how it is necessary to pick the correct solution among those found. We have also shown that constructing the distribution using a limited number of moments is of limited precision, and the result has to be understood as an approximation; for example, one can observe $\rho <0$ due to this inaccuracy. Careful handling, however, still allows it to be used for obtaining valuable insight into the model.  

\subsubsection*{Acknoweledgement}

This research was supported by VEGA 1/0025/23 grant \emph{Matrix models and quantum gravity}. The authors would like to acknowledge the contribution of the COST Action CA23130, \textit{Bridging high and low energies in search of quantum gravity} and COST Action CA21109, \textit{Cartan geometry, Lie, Integrable Systems, quantum group Theories for Applications}; parts of this work were disseminated during Cost Action CaLISTA General Meeting 2025 at Corfu Summer Institute. The authors would also like to thank Juraj Tekel for his valuable comments and suggestions. 

\appendix
\section{Importance of the initial distribution}
\label{sec:ImportanceoftheInitialDistribution}
In applying the approximation method (\ref{eq:aproximacia}), we found that the choice of the initial distribution \(P(x)\) strongly influences convergence and overall accuracy. Two aspects proved especially important: the normalization interval and the functional form of the initial distribution.

The support of \(P(x)\) should approximately match that of the target distribution. If the interval is too narrow, the method diverges quickly and fails to approximate the desired density even for small values of \(M\). If it is too wide, the approximation still converges but requires larger \(M\), which can lead to numerical complications.

\begin{figure}[H]
    \centering
    \begin{minipage}{0.57\textwidth}
        \includegraphics[width=\textwidth]{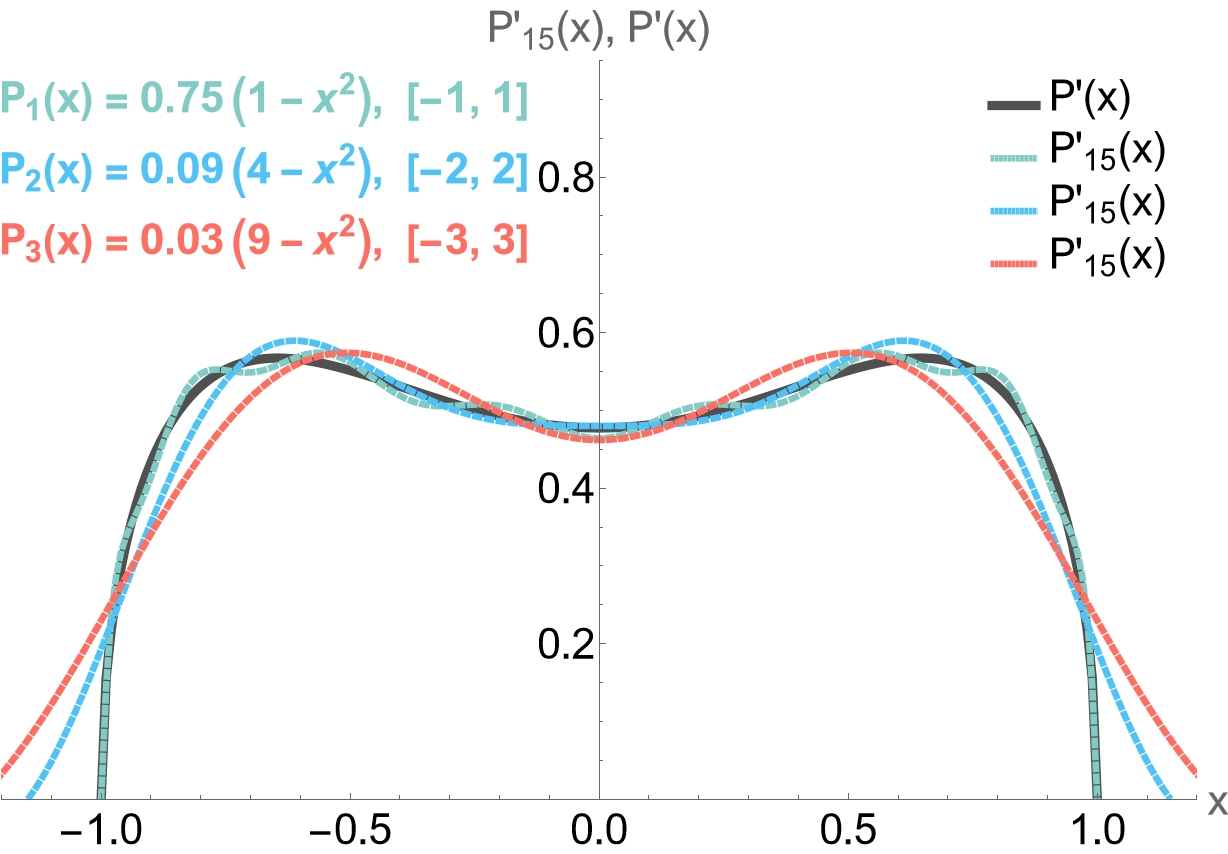}
    \end{minipage}
    \caption[Initial function –– choice of interval]{Comparison of the approximation \(P'_{15}(x)\) with the target density \(P'(x)\) for different initial functions and normalization intervals.}
    \label{fig:IFinterval}
\end{figure}

Fig.\ref{fig:IFinterval} compare three initial functions -- parabolic, constant, and Gaussian --normalized on intervals \([-1,1]\), \([-2,2]\), and \([-3,3]\). The target density \(P'(x)\) corresponds to the one-cut solution (\ref{eq:onecut}) for \(r=1\) and \(g=1\), with the support over the interval \([-1,1]\), shown as the solid black line.
\begin{figure}[H]
    \centering
    \begin{minipage}{0.48\textwidth}
        \includegraphics[width=\textwidth]{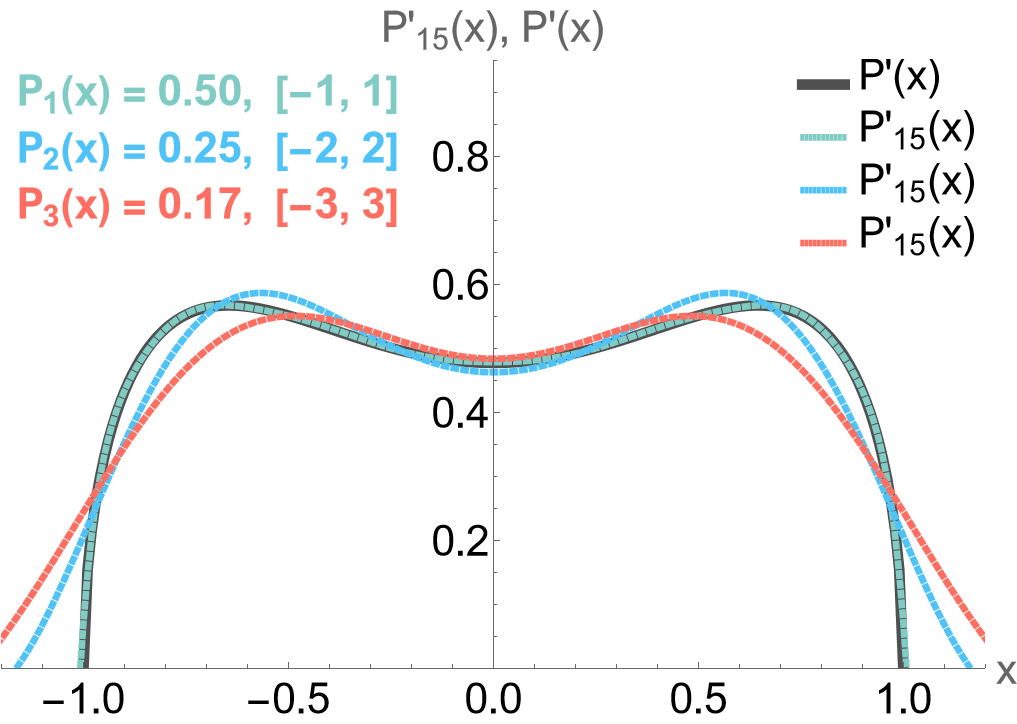}
    \end{minipage}
    \hspace{0.3cm}
    \begin{minipage}{0.48\textwidth}
        \includegraphics[width=\textwidth]{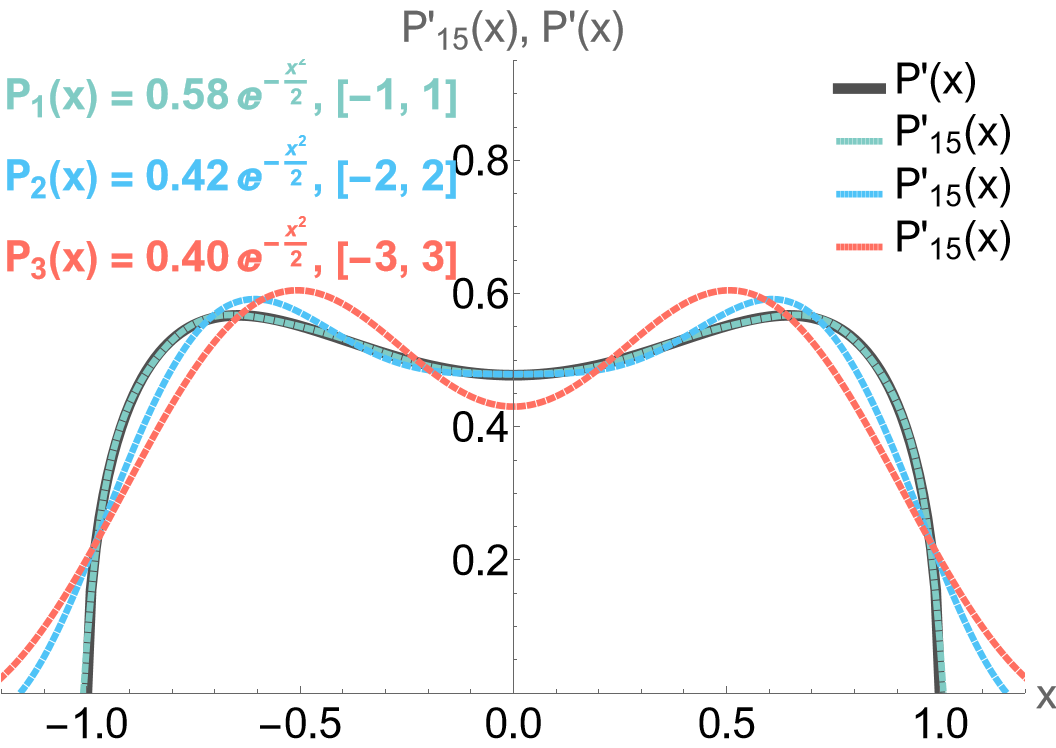}
    \end{minipage}
\end{figure}

The quadratic error values in Tab.~\ref{tab:L2error} quantify this effect:

\begin{table}[H]
    \centering
    \begin{tabular}{|c|ccc|}
        \hline
        Interval & Parabolic & Constant & Gaussian \\
        \hline
        \( [-1,1] \) & 0.00026 & 0.00001 & 0.00001 \\
        \( [-2,2] \) & 0.00189 & 0.00257   & 0.00209   \\
        \( [-3,3] \) & 0.00447 & 0.00508   & 0.00678   \\
        \hline
    \end{tabular}
    \caption{Quadratic error of the approximation for different intervals and initial distributions.}
    \label{tab:L2error}
\end{table}

These results clearly show that using the correct interval significantly improves convergence. Constant and Gaussian functions perform equally well in this respect, whereas the parabolic function is less stable.

\begin{figure}[H]
    \centering
    \begin{minipage}{0.48\textwidth}
        \includegraphics[width=\textwidth]{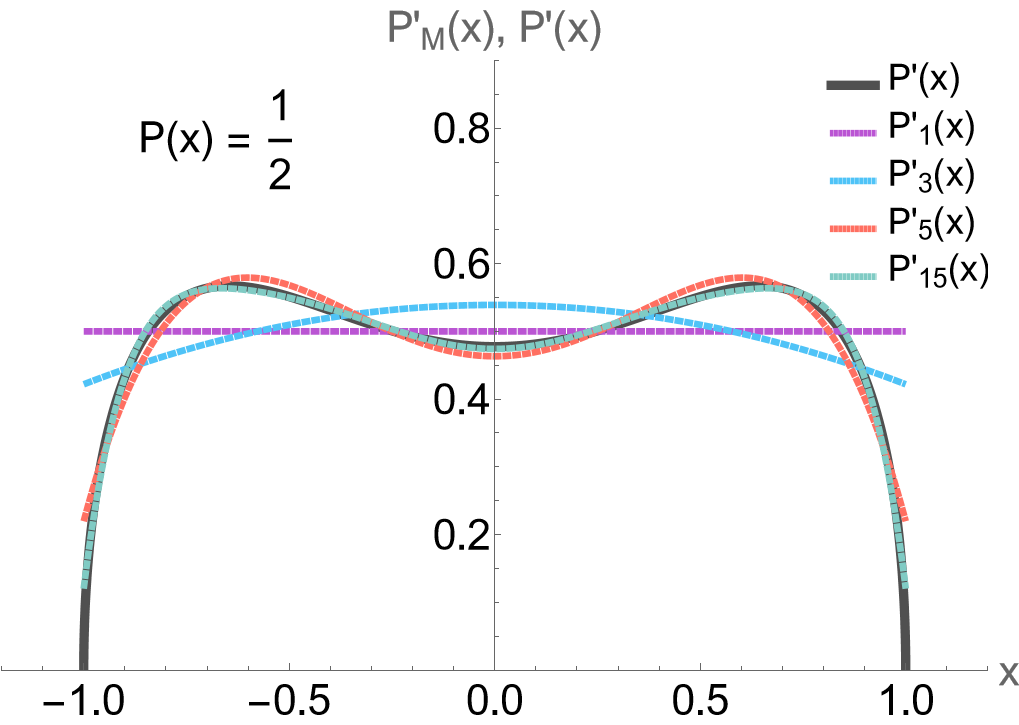}
    \end{minipage}
    \hspace{0.3cm}
    \begin{minipage}{0.48\textwidth}
        \includegraphics[width=\textwidth]{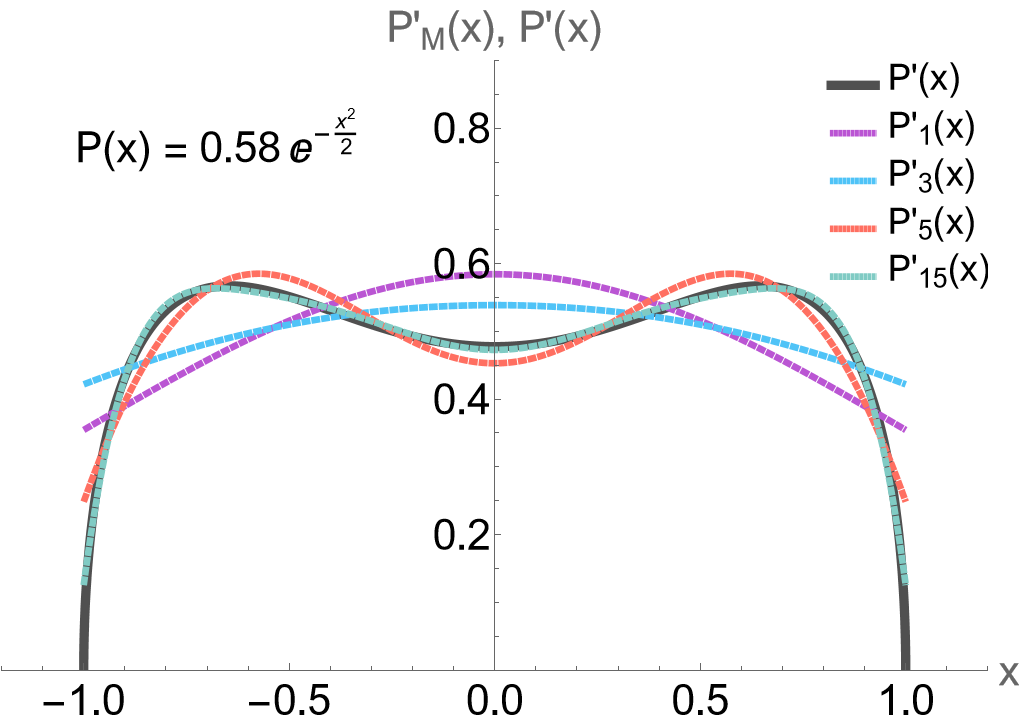}
    \end{minipage}
    \centering
    \begin{minipage}{0.5\textwidth}
        \includegraphics[width=\textwidth]{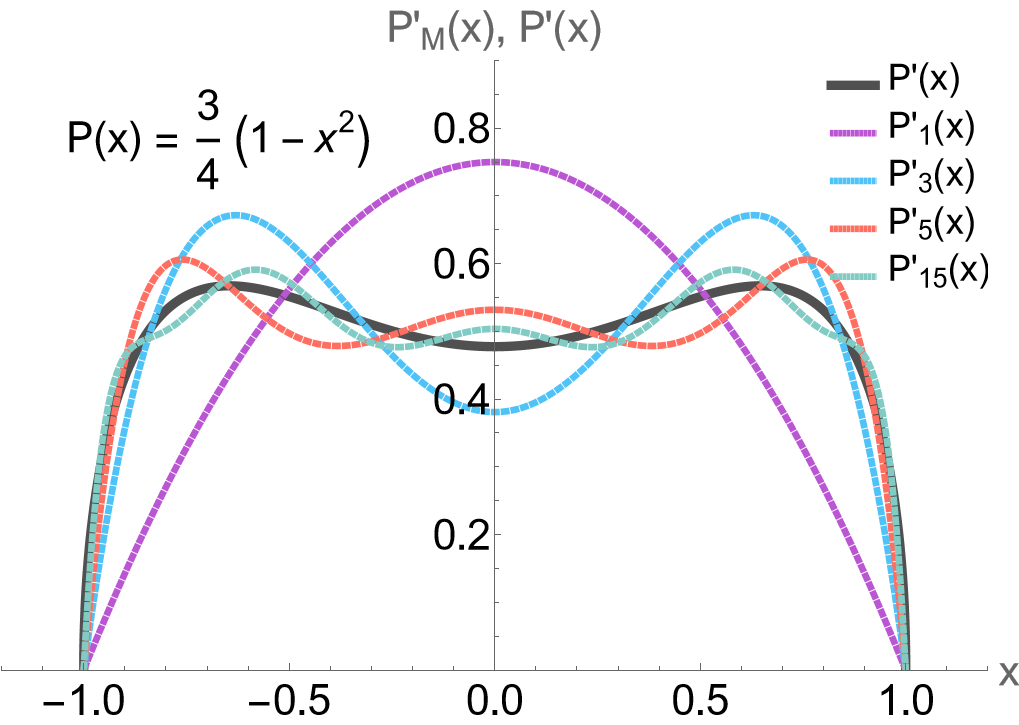}
    \end{minipage}
    \caption[Initial function -- type of distribution]{Approximation \(P'_M(x)\) compared with \(P'(x)\) for various \(M\) and initial functions.}
    \label{fig:IFtypeM}
\end{figure}

With an appropriate interval, the type of initial distribution also affects convergence. Fig.\ref{fig:IFtypeM} illustrates how the initial distribution functions behave for \(M = 1,3,5,15\).

\begin{figure}[H]
    \centering
    \begin{minipage}{0.51\textwidth}
        \includegraphics[width=\textwidth]{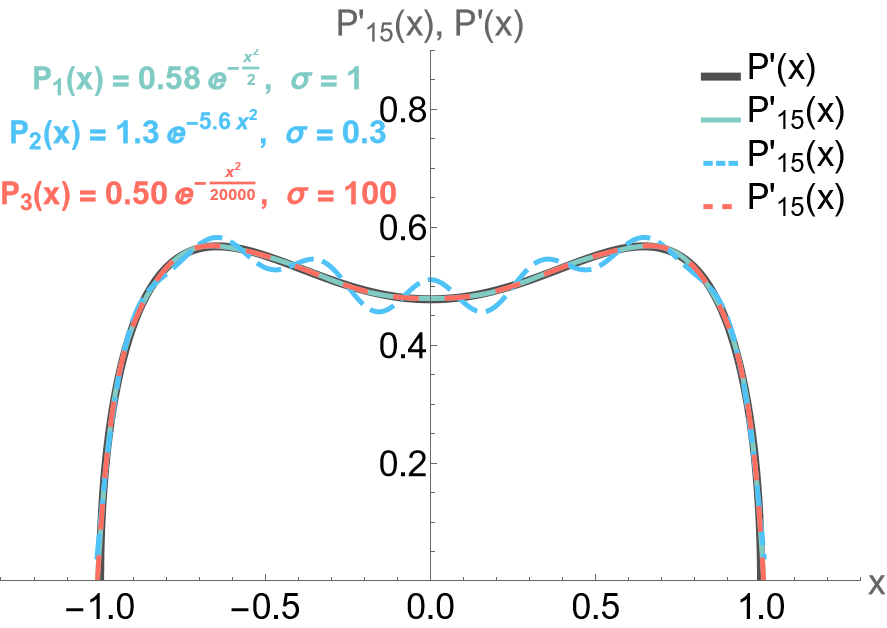}
    \end{minipage}
    \caption[Initial function –– influence of \(\sigma\)]{Influence of the Gaussian parameter \(\sigma\) on approximation accuracy.}
    \label{fig:IFsigma}
\end{figure}

Constant and Gaussian distributions converge rapidly toward \(P'(x)\) even for small \(M\), while the parabolic choice exhibits persistent oscillations and slower convergence. For Gaussian initial distributions, the parameter \(\sigma\) influences accuracy, see Fig.~\ref{fig:IFsigma}. Values of \(\sigma \geq 1\) yield significantly better results than smaller \(\sigma.\)

In our calculations, a constant initial distribution was typically sufficient: its convergence properties were comparable to the Gaussian case while avoiding the need to tune \(\sigma\). Moreover, when the target distribution is asymmetric, choosing a constant initial distribution becomes even more effective, as it tends to converge faster than a Gaussian one. Nonetheless, these comparisons highlight that both interval and functional form must be chosen carefully to ensure stability and accuracy of the approximation method.

\section{Numerical details}
\label{sec:NumericalDetails}

Our calculations were performed using Python and Wolfram Mathematica. The code is structured into three main parts:

\begin{enumerate}
    \item \textbf{Bootstrap Method:} This part generates the moments required for the approximation method. The bootstrap procedure is sensitive to several factors, notably the size of the bootstrap matrix. For symmetric solutions, the relevant matrix sizes are shown in Fig.\ref{fig:m1QP}. Increasing the matrix size requires a corresponding reduction in the iterative step used for searching for the desired moments. For symmetric solutions, where only the second moment is needed, a smaller precision is sufficient. In contrast, when considering also asymmetric solutions, iteration over both \(m_1\) and \(m_2\) moments is required, demanding higher precision and smaller iterative steps. The matrix sizes needed for good convergence of \(m_1\) in asymmetric cases are shown in Fig.\ref{fig:CAMMMm1forR}.

    \item \textbf{Approximation Method:} The second part computes the approximate probability distribution. Here, the choice of the \textit{initial distribution} and its normalized interval plays a crucial role. For symmetric solutions and one-cut asymmetric cases, we chose symmetric intervals. Typical parameters were \(M=10-15\) for symmetric solutions and \(M \approx 8\) for asymmetric one-cut solutions. It is worth noting that when a larger interval is chosen, one might expect that \(M\) could be increased correspondingly. However, in practice, convergence sometimes fails. The issue arises due to the normalization in the relevant formula: for large \(M\), the method involves dividing by small numbers, which can render the solution numerically unstable.

    \item \textbf{Free Energy Calculation:} Once the probability density is obtained, its free energy is computed. This process is repeated for various values of \(r\) and \(g\), separately for symmetric and asymmetric solutions. The preferred solutions, which are then visualized in a phase diagram, are determined by comparing free energies. 
\end{enumerate}

To distinguish between one-cut and two-cut solutions, we applied simple threshold-based criteria. For symmetric solutions, a distribution was classified as two-cut if its value at \(x=0\) remained below 0.05; otherwise, it was considered one-cut. For asymmetric solutions, a distribution was classified as two-cut if it exceeded 0.01 on both sides, otherwise it was considered one-cut.

\bibliographystyle{unsrt} 
\bibliography{refs}       

\begin{thebibliography}{10}

\bibitem{Eynard:2015aea}
Bertrand Eynard, Taro Kimura, and Sylvain Ribault.
\newblock Random matrices, 2015.
\newblock arXiv:1510.04430 [math-ph].

\bibitem{Jha:2021exo}
Rohan~G. Jha.
\newblock Introduction to monte carlo for matrix models.
\newblock {\em SciPost Phys. Lect. Notes}, 46:1, 2022.

\bibitem{Han:2020bkb}
Xizhi Han, Sean~A. Hartnoll, and Jorrit Kruthoff.
\newblock Bootstrapping matrix quantum mechanics.
\newblock {\em Phys. Rev. Lett.}, 125(4):041601, 2020.

\bibitem{Lin:2020mme}
Henry~W. Lin.
\newblock {Bootstraps to strings: solving random matrix models with positivity}.
\newblock {\em JHEP}, 06:090, 2020.

\bibitem{Khalkhali:2020jzr}
Masoud Khalkhali, Nathan Pagliaroli, Andrei Parfeni, and Brayden Smith.
\newblock {Bootstrapping the critical behavior of multi-matrix models}.
\newblock {\em JHEP}, 25:158, 2020.

\bibitem{Hessam:2021byc}
Hamed Hessam, Masoud Khalkhali, and Nathan Pagliaroli.
\newblock {Bootstrapping Dirac ensembles}.
\newblock {\em J. Phys. A}, 55(33):335204, 2022.

\bibitem{Kazakov:2021lel}
Vladimir Kazakov and Zechuan Zheng.
\newblock {Analytic and numerical bootstrap for one-matrix model and {\textquotedblleft}unsolvable{\textquotedblright} two-matrix model}.
\newblock {\em JHEP}, 06:030, 2022.

\bibitem{Aikawa:2021qbl}
Yu~Aikawa, Takeshi Morita, and Kota Yoshimura.
\newblock {Bootstrap method in harmonic oscillator}.
\newblock {\em Phys. Lett. B}, 833:137305, 2022.

\bibitem{Berenstein:2021loy}
David Berenstein and George Hulsey.
\newblock {Bootstrapping more QM systems}.
\newblock {\em J. Phys. A}, 55(27):275304, 2022.

\bibitem{Bhattacharya:2021btd}
Jyotirmoy Bhattacharya, Diptarka Das, Sayan~Kumar Das, Ankit~Kumar Jha, and Moulindu Kundu.
\newblock {Numerical bootstrap in quantum mechanics}.
\newblock {\em Phys. Lett. B}, 823:136785, 2021.

\bibitem{Zheng:2023bjj}
Zechuan Zheng.
\newblock {\em {Bootstrap Method in Theoretical Physics}}.
\newblock PhD thesis, Ecole Normale Sup{\'e}rieure, Paris, 2023.

\bibitem{Khalkhali:2023onm}
Masoud Khalkhali and Nathan Pagliaroli.
\newblock {Coloured combinatorial maps and quartic bi-tracial 2-matrix ensembles from noncommutative geometry}.
\newblock {\em JHEP}, 05:186, 2024.

\bibitem{Kazakov:2024ool}
Vladimir Kazakov and Zechuan Zheng.
\newblock {Bootstrap for finite N lattice Yang-Mills theory}.
\newblock {\em JHEP}, 03:099, 2025.

\bibitem{Lawrence:2024mnj}
Scott Lawrence, Brian McPeak, and Duff Neill.
\newblock {Bootstrapping time-evolution in quantum mechanics}.
\newblock 12 2024.

\bibitem{Huang:2025sua}
Zhijian Huang and Wenliang Li.
\newblock {Bootstrapping periodic quantum systems}.
\newblock 7 2025.

\bibitem{Blacker_2022}
Matthew~J. Blacker, Arpan Bhattacharyya, and Aritra Banerjee.
\newblock Bootstrapping the kronig-penney model.
\newblock {\em Physical Review D}, 106(11), December 2022.

\bibitem{Berenstein:2021dyf}
David Berenstein and George Hulsey.
\newblock {Bootstrapping Simple QM Systems}.
\newblock 8 2021.

\bibitem{Berenstein:2022ygg}
David Berenstein and George Hulsey.
\newblock {Anomalous bootstrap on the half-line}.
\newblock {\em Phys. Rev. D}, 106(4):045029, 2022.

\bibitem{Khan:2022uyz}
Sakil Khan, Yuv Agarwal, Devjyoti Tripathy, and Sachin Jain.
\newblock {Bootstrapping PT symmetric quantum mechanics}.
\newblock {\em Phys. Lett. B}, 834:137445, 2022.

\bibitem{Nakayama:2022ahr}
Yu~Nakayama.
\newblock {Bootstrapping microcanonical ensemble in classical system}.
\newblock {\em Mod. Phys. Lett. A}, 37(09):2250054, 2022.

\bibitem{Tchoumakov:2021mnh}
Serguei Tchoumakov and Serge Florens.
\newblock {Bootstrapping Bloch bands}.
\newblock {\em J. Phys. A}, 55(1):015203, 2022.

\bibitem{RevModPhys.69.731}
C.~W.~J. Beenakker.
\newblock Random-matrix theory of quantum transport.
\newblock {\em Rev. Mod. Phys.}, 69:731--808, Jul 1997.

\bibitem{Szabo:2001kg}
Richard~J. Szabo.
\newblock {Quantum field theory on noncommutative spaces}.
\newblock {\em Phys. Rept.}, 378:207--299, 2003.

\bibitem{Karabali:2006eg}
Dimitra Karabali and V.~P. Nair.
\newblock {Quantum Hall effect in higher dimensions, matrix models and fuzzy geometry}.
\newblock {\em J. Phys. A}, 39:12735--12764, 2006.

\bibitem{Balachandran:2005ew}
A.~P. Balachandran, S.~Kurkcuoglu, and S.~Vaidya.
\newblock {Lectures on fuzzy and fuzzy SUSY physics}.
\newblock 11 2005.

\bibitem{Ydri:2016dmy}
Badis Ydri.
\newblock {\em {Lectures on Matrix Field Theory}}, volume 929.
\newblock Springer, 2017.

\bibitem{Tekel:2015uza}
Juraj Tekel.
\newblock {Phase strucutre of fuzzy field theories and multitrace matrix models}.
\newblock {\em Acta Phys. Slov.}, 65(5):369--468, 2015.

\bibitem{Subjakova:2020prh}
M{\'a}ria {\v{S}}ubjakov{\'a} and Juraj Tekel.
\newblock {Fuzzy field theories and related matrix models}.
\newblock {\em PoS}, CORFU2019:189, 2020.

\bibitem{tekel2012constructing}
Juraj Tekel and L~Cohen.
\newblock Constructing and estimating probability distributions from moments.
\newblock In {\em Automatic target recognition XXII}, volume 8391, pages 114--123. SPIE, 2012.

\bibitem{bukor2024simple}
Benedek Bukor and Juraj Tekel.
\newblock Cubic asymmetric multitrace matrix model.
\newblock {\em Journal of Physics A: Mathematical and Theoretical}, 58(25):255203, jun 2025.

\bibitem{Livan_2018}
Giacomo Livan, Marcel Novaes, and Pierpaolo Vivo.
\newblock {\em Introduction to Random Matrices}.
\newblock Springer International Publishing, 2018.

\bibitem{szego75}
G\'{a}bor Szeg\"{o}.
\newblock {\em Orthogonal Polynomials}.
\newblock American Mathematical Society, Providence, RI, 1975.

\bibitem{simon1999classicalmomentproblemselfadjoint}
Barry Simon.
\newblock The classical moment problem as a self-adjoint finite difference operator, 1999.

\end{thebibliography}

\end{document}